\documentclass[letterpaper, 10pt, conference]{ieeeconf}  

\IEEEoverridecommandlockouts                              

\usepackage{circledsteps} 

\definecolor{QdaiColor}{HTML}{93023B} 
\pgfkeys{/csteps/inner color=white}
\pgfkeys{/csteps/outer color=white}
\pgfkeys{/csteps/fill color=QdaiColor}
\pgfkeys{/csteps/inner ysep=7pt}
\pgfkeys{/csteps/inner xsep=4pt}
\usepackage{graphicx}
\usepackage{subfigure}
\usepackage{multirow}

\usepackage{comment} 

\let\labelindent\relax 
\usepackage{enumitem}

\newcommand{\email}[1]{{\footnotesize{#1}}}

\usepackage[colorlinks=true, linkcolor=black, urlcolor=black, citecolor=black]{hyperref}

\usepackage{tikz} 
\newcommand\copyrighttext{%
  \normalfont \tiny \textcopyright 2025 IEEE. Personal use of this material is permitted. Permission from IEEE must be obtained for all other uses, in any current or future media, including reprinting/republishing this material for advertising or promotional purposes, creating new collective works, for resale or redistribution to servers or lists, or reuse of any copyrighted component of this work in other works.\\
  Cite this article as follows: M. Ozaki, L. Chen, S. Naganuma, V. Švábenský, F. Okubo, A. Shimada. \textit{PALM: PAnoramic Learning Map Integrating Learning Analytics and Curriculum Map for Scalable Insights Across Courses}. In Proceedings of the 2025 IEEE International Conference on Systems, Man, and Cybernetics (SMC '25). Vienna, Austria, 2025. DOI: \href{https://doi.org/10.1109/SMC58881.2025.11343513}{10.1109/SMC58881.2025.11343513}.}
\newcommand\copyrightnotice{%
\begin{tikzpicture}[remember picture,overlay]
\node[anchor=south,yshift=5pt] at (current page.south) {\fbox{\parbox{\dimexpr\textwidth-\fboxsep-\fboxrule\relax}{\copyrighttext}}};
\end{tikzpicture}%
}

\title{\LARGE \bf
PALM: PAnoramic Learning Map Integrating Learning Analytics \\
and Curriculum Map for Scalable Insights Across Courses
\copyrightnotice
}

\author{
    Mahiro Ozaki$^{1}$,~~Li Chen$^{2}$,~~Shotaro Naganuma$^{3}$, \\%
    Valdemar Švábenský$^{4}$,~~Fumiya Okubo$^{5}$,~~and~~Atsushi Shimada$^{5}$%
    \thanks{
        $^{1}$\,Graduate School of Information Science and Electrical Engineering, Kyushu University, Japan.
        \email{ozaki.mahiro.493@\allowbreak s.kyushu-u.ac.jp},
        $^{2}$\,Division of Math, Sciences, and Information Technology in Education, Osaka Kyoiku University, Japan.
        \email{chen-l68@\allowbreak cc.osaka-kyoiku.ac.jp},
        $^{3}$\,Promoting Organization for Future Creators, Kyushu University, Japan.
        \email{naganuma.shotaro.062@\allowbreak m.kyushu-u.ac.jp},
        $^{4}$\,Faculty of Informatics, Masaryk University, Czech Republic.
        \email{valdemar@\allowbreak mail.muni.cz},
        $^{5}$\,Faculty of Information Science and Electrical Engineering, Kyushu University, Japan.
        \email{fokubo@\allowbreak ait.kyushu-u.ac.jp}, 
        \email{atsushi@\allowbreak ait.kyushu-u.ac.jp}
    }
}

\begin{document}

\maketitle
\thispagestyle{empty}
\pagestyle{empty}

\begin{abstract}
This study proposes and evaluates the \textit{PAnoramic Learning Map (PALM)}, a learning analytics (LA) dashboard designed to address the scalability challenges of LA by integrating curriculum-level information. Traditional LA research has predominantly focused on individual courses or learners and often lacks a framework that considers the relationships between courses and the long-term trajectory of learning. To bridge this gap, PALM was developed to integrate multilayered educational data into a curriculum map, enabling learners to intuitively understand their learning records and academic progression. We conducted a system evaluation to assess PALM's effectiveness in two key areas: (1) its impact on students’ awareness of their learning behaviors, and (2) its comparative performance against existing systems. The results indicate that PALM enhances learners' awareness of study planning and reflection, particularly by improving \textit{perceived behavioral control} through the visual presentation of individual learning histories and statistical trends, which clarify the links between learning actions and outcomes. Although PALM requires ongoing refinement as a system, it received significantly higher evaluations than existing systems in terms of visual appeal and usability. By serving as an information resource with previously inaccessible insights, PALM enhances self-regulated learning and engagement, representing a significant step beyond conventional LA toward a comprehensive and scalable approach.
\end{abstract}


\section{INTRODUCTION}
Learning analytics (LA) is an academic field that aims to optimize educational activities through the collection and analysis of educational data, and the provision of feedback. 
In recent years, LA has been increasingly recognized as a means of supporting self-regulated learning (SRL). 
SRL is a prominent learning theory that emphasizes learners' ability to actively plan, monitor, and regulate their learning processes~\cite{zimmerman1998developing}, enabling them to adjust their learning strategies based on real-time data and insights~\cite{azevedo2022lessons}.
To this end, LA research integrates knowledge from diverse disciplines such as computer science, education, data mining, statistics, and behavioral sciences, and has developed various tools to enhance learner autonomy.

Furthermore, in recent LA research, the perspective of scalability has gained increasing attention. There is a growing demand for LA systems that can handle large-scale data and support diverse learners and environments. 
For instance, Lonn et al.~\cite{Lonn2023issues} developed an infrastructure with a university IT department to enable real-time support across an entire institution. 
Additionally, Ruip\'{e}rez-Valiente et al.~\cite{ruiperez2016scaling} proposed a system for visualizing and analyzing learning data in massive open online courses, where each course may have over 100,000 learners.
These efforts highlight significant progress toward scalable implementations of LA across varied educational settings.
In parallel, LA dashboards have been developed to support SRL, course recommendation, and competency tracking, showing positive impacts on student performance~\cite{kimJ2016, santos2012empowering, Silvola2023learning, grann2014competency}. 

However, most of these studies primarily focus on individual courses or learners, often relying on short-term learning data. 
As a result, their applicability to long-term support and curriculum-level integration remains limited.
University education is inherently designed as an integrated curriculum rather than as an isolated individual course~\cite{roehrig2021understanding}. 
For example, foundational knowledge in mathematics influences students’ understanding of physics and statistics, demonstrating that different courses are interconnected and shape learning outcomes. 
Without feedback that accounts for such cross-course interdependencies, support for learners is likely to remain fragmented, making it difficult for LA to achieve its essential goal of providing comprehensive learning support.

\begin{figure}[t]
    \centering
    \includegraphics[width=0.44\textwidth]{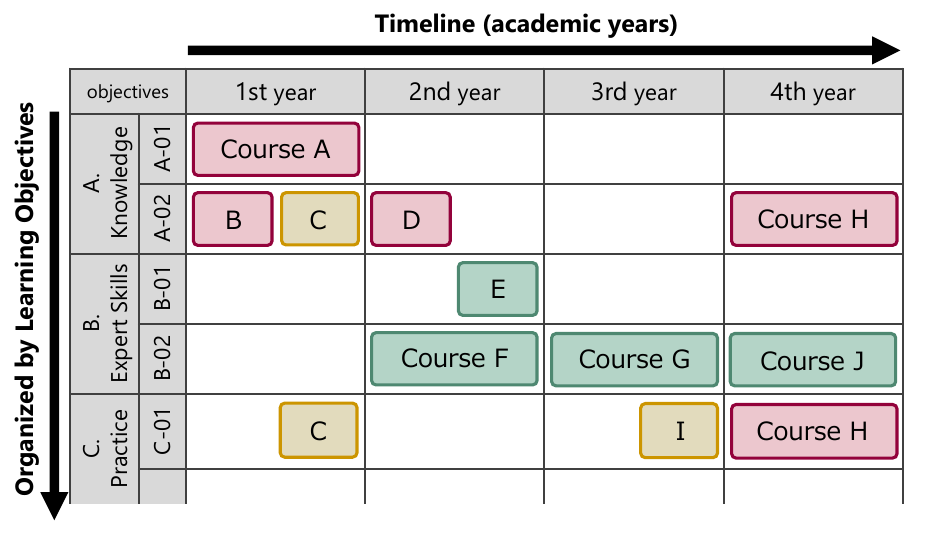} 
    \caption{Illustration of the curriculum map overview}
    \label{fig:curriculum_map}
\end{figure}

Therefore, this study focuses on \textit{curriculum mapping} as a scaled educational process encompassing multiple courses and proposes and implements an LA dashboard to support SRL.
In the field of educational management, many institutions have adopted \textit{curriculum maps} created based on curriculum policies \cite{originalCurriculumMap, instructionalCurriculumMapping}. As shown in Figure~\ref{fig:curriculum_map}, curriculum maps provide a visual representation of the curriculum structure aligned with learning objectives on the vertical axis and a timeline (e.g., academic years or semesters) on the horizontal axis. 
These maps are used to support curriculum design and review, and offer an overview of learning content from enrollment to graduation~\cite{michael2019application, plaza2007curriculum}.

Although curriculum maps were originally designed for educational management, increasing emphasis has been placed on supporting learners’ SRL at the curriculum level. Consequently, recent studies explored the potential of using curriculum maps as tools for learners to independently plan and reflect on their learning~\cite{Rawle2017curriculm}. However, simply providing learners with a curriculum map is insufficient for achieving intuitive and comprehensive learning support, as noted in previous studies \cite{livedUpToHype?}. 
The following factors contribute to this challenge.
\begin{itemize} 
    \item Curriculum maps do not explicitly show detailed learning content, course interrelations, or dependencies in learning progression. 
    \item Students unfamiliar with the curriculum’s structure or intent may struggle to interpret and use the map effectively. 
    \item Curriculum maps lack linkage to individual learning outcomes, limiting integration with LA dashboards and syllabus systems.
\end{itemize}

Furthermore, many studies in this field have concluded with technical validation or the design of LA dashboards and have not progressed to system implementation or providing feedback to students. Although research on user interface design, system architecture, and data processing has highlighted the potential of LA in educational support, studies presenting systems that are directly usable by learners remain limited \cite{Kang20219database, Schutte2018using, Hotta2023construction}. 
As a result, while the theoretical usefulness of LA at the curriculum level has been discussed, efforts to translate these findings into practical applications remain another underexplored challenge.

To address these challenges, this study proposes a system \textit{\textbf{PAnoramic Learning Map (PALM)}}, which integrates micro-level LA with macro-level curriculum maps. PALM inherits the visibility features of curriculum maps and supports an intuitive understanding of course-related learning objectives based on spatial proximity. In addition, PALM overlays multilayered map information representing learning activities and outcomes related to individual courses and through LA on the same map. 
Through this approach, the digitalized curriculum map functions as a \textit{hub} that integrates dispersed information from existing systems (e.g., Learning Management System, syllabi, and traditional curriculum maps), enabling a novel form of educational support by systematically presenting interconnected learning information.

PALM aims to support stakeholders, such as curriculum designers, educators, and students, in linking the overall curriculum structure with practical educational activities on a larger scale. 
This support includes curriculum design, teaching methods, learning activities, planning, reflection, and continuous improvement across multiple courses. 
From a broader perspective, PALM enables the integration and effective use of cross-curricular learning data by extending the traditional scope of LA, which primarily focuses on individual courses and learners. 
Through this approach, PALM makes a comprehensive and scalable contribution to enhancing learners' SRL (planning and reflection) and engagement.
To the best of our knowledge, no previous research has implemented such a system as a concrete application or evaluated it with users.
Based on the above, the following points were set as the research questions:

\begin{enumerate}[leftmargin=2.4em, itemindent=0pt] 
    \item[\textbf{RQ1}] How does PALM influence students' attitudes and intentions regarding their self-regulated learning behaviors, such as study planning and reflection?
    \item[\textbf{RQ2}] How do students evaluate PALM as a LA dashboard, particularly its curriculum map interface incorporating multilayered LA information, in comparison with existing systems (e.g., LMS, syllabi, traditional curriculum maps)?
\end{enumerate}
\noindent
By addressing these research questions, this study examines the role of PALM as a LA dashboard and provides insight into changes in learners’ behavioral intentions and system evaluations. 
Ultimately, this study aims to contribute to providing long-term learning support across the entire curriculum.

\begin{figure}[t]
    \centering
    \includegraphics[width=0.51\textwidth]{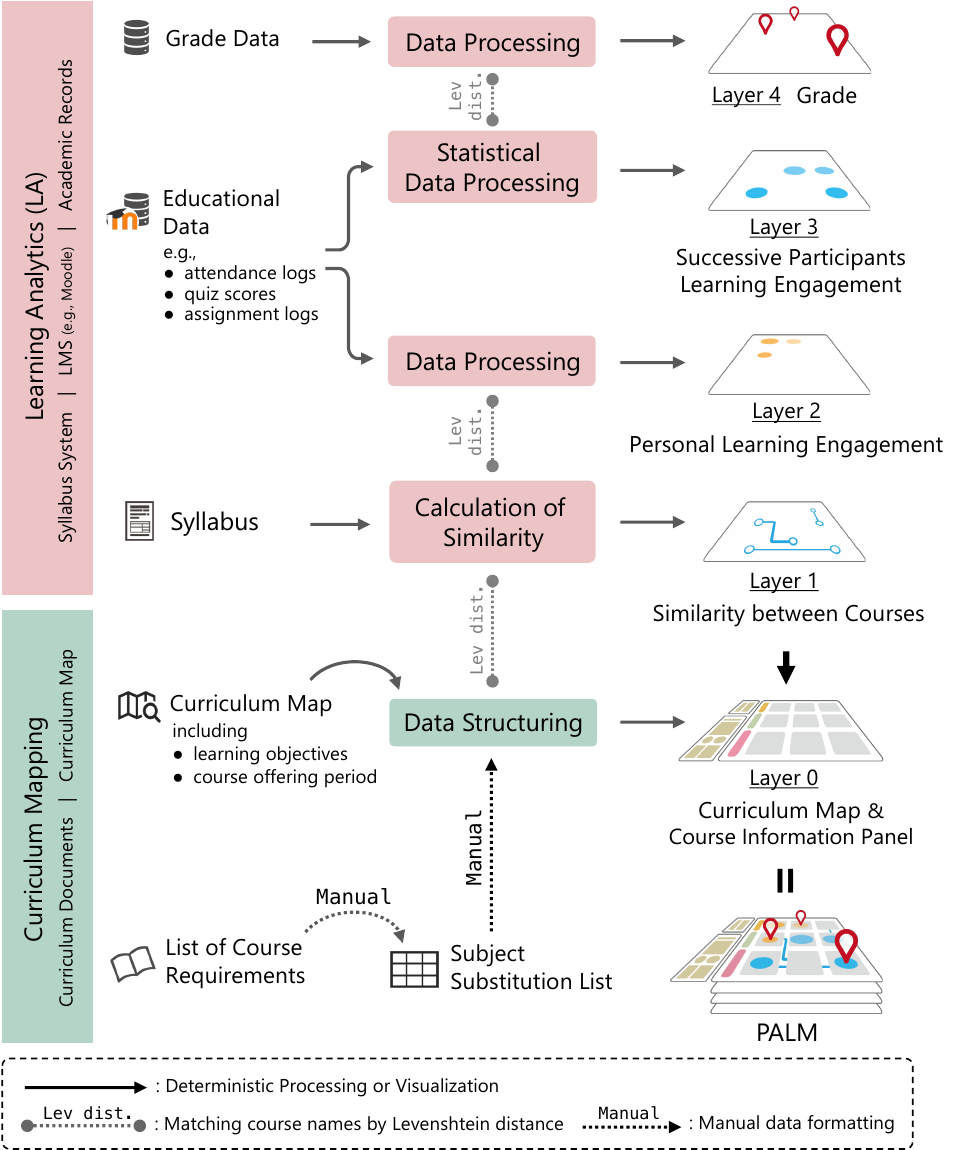} 
    \caption{Design concept of PALM}
    \label{fig:design_concept}
\end{figure}

\section{PROPOSED SYSTEM: ``PALM''}
\label{sec:proposed_system}

\begin{figure*}[t]
    \centering
    \includegraphics[width=1.0\textwidth]{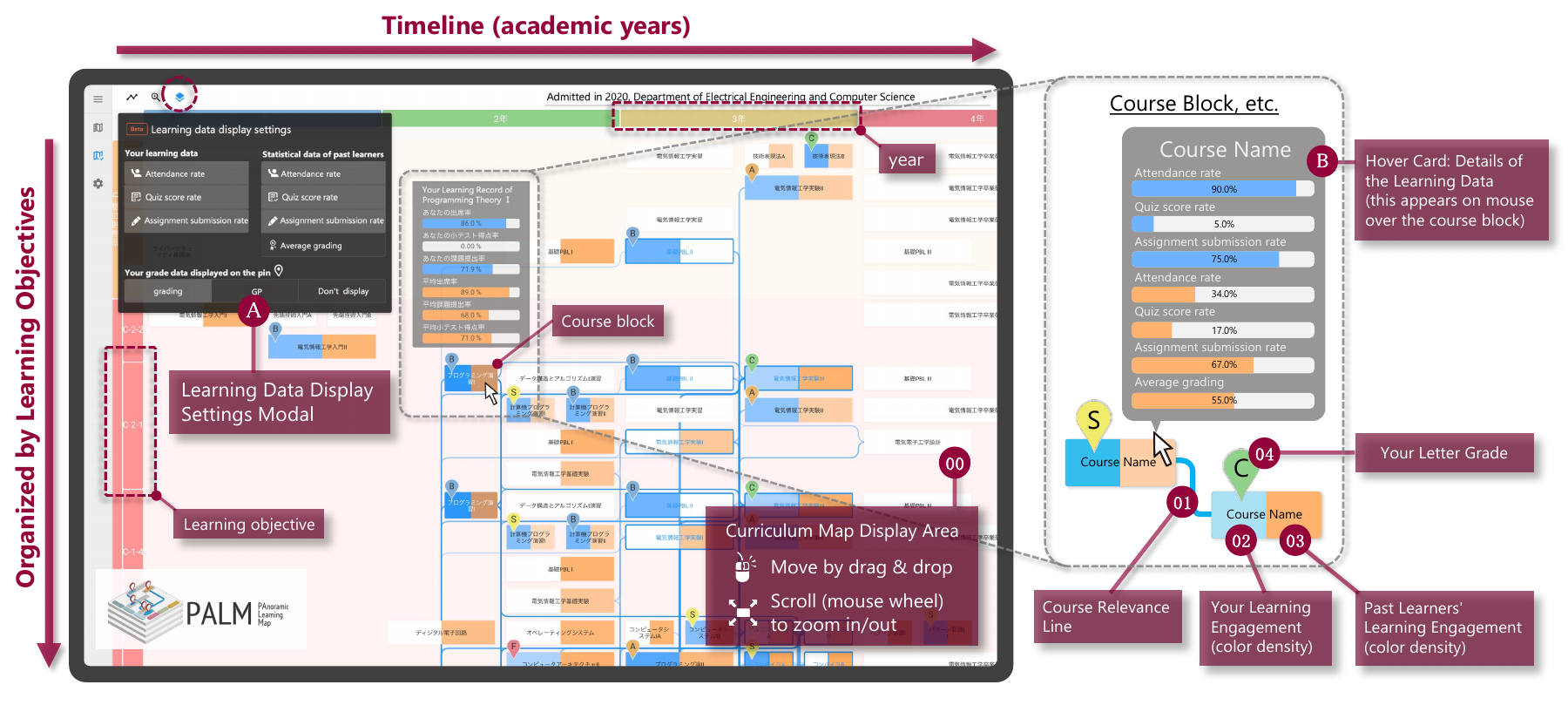} 
    \caption{User interface of PALM}
    \label{fig:PALM_UI}
\end{figure*}

\subsection{Design Concept}
\label{sec:design_comcept}

This section outlines the design concept of the proposed system, \textit{PAnoramic Learning Map (PALM)}. 
The objective of PALM is to implement a representation method that links educational support information from the perspective of LA to curriculum maps that serve as \textit{hubs} for educational management. 
PALM is inspired by geographic information systems (GIS), which overlay various types of information as layers on a base map to visualize correlations and trends among map elements. 
In this analogy, as shown in Figure~\ref{fig:design_concept}, the base map corresponds to the curriculum map, whereas the information layers represent the data derived from LA (collected through existing systems). 
Users can customize the types of data displayed on the curriculum map to achieve visualizations tailored to their specific purposes.

Figure~\ref{fig:design_concept} illustrates the various layers and the corresponding data sources required to construct them.
As arranged along the pink and green rectangles, these data types were traditionally handled from the perspectives of LA and curriculum mapping, respectively. 
After undergoing the respective processing steps described in the central column, these data are ultimately visualized in the form of layers 0 through 4, as shown on the right side, which constitute the user interface of PALM.

\subsection{User Interface}
\label{sec:user_interface}

Figure~\ref{fig:PALM_UI} illustrates the PALM user interface and a detailed view focusing on a \textit{course block}. 
The features are introduced below.

\begin{enumerate}[leftmargin=1.8em, itemindent=0pt]
    \item[\Circled{00}] 
    \textit{\textbf{Curriculum Map Display Area (Layer 0):}}
    All courses in the selected curriculum (for example, numbering around 180 for each faculty and program at Kyushu University) are shown as \textit{course blocks} in a two-dimensional layout, similar to the original curriculum map. 
    This corresponds to the ``base map'' concept in GIS. 
    The map can be freely navigated using drag-and-drop for movement and a mouse wheel for zooming.
    \item[\Circled{01}] 
    \textit{\textbf{Course Relevance Lines (Layer 1):}}
    The relationships between courses are visualized using blue lines, with thicker lines indicating stronger connections between course blocks. Textual data such as \textit{course overview} and \textit{lecture plan} from syllabi were vectorized using TF-IDF~\cite{salton1988tfidf}, and cosine similarity was computed to visualize these relationships. The resulting visualizations were found to be reasonably valid by learners (see Section~\ref{sec:datacollection_analysys_system} and \ref{sec:result_system}).
    \item[\Circled{02}] 
    \textit{\textbf{Individual Learning Engagement (Layer 2):}}
    The left half of each course block is overlaid with a blue-shaded area representing individual learning engagement, which is defined as the average of attendance rates, quiz scores, and assignment submissions. 
    The combination of displayed information can be freely customized through \textit{display settings modal} ({\footnotesize \Circled{A}} in Figure~\ref{fig:PALM_UI}).
    \item[\Circled{03}] 
    \textit{\textbf{Past Learners’ Engagement (Layer 3):}}
    The right half of each course block is overlaid with an orange-shaded area representing the learning engagement of past course takers. Other settings follow those of the individual engagement.
    \item[\Circled{B}] 
    \textit{\textbf{Hover Card of Details:}}
    Hovering the mouse over a course block displays detailed information about each learning data element in a popup.
    \item[\Circled{04}] 
    \textit{\textbf{Letter Grade (Layer 4):}}
    Grade information is displayed as pin-shaped markers on the map. The display settings modal dialog described later allows switching between letter grades, grade point, and no display.
\end{enumerate}

\section{EVALUATION METHOD}
\label{sec:evalation}
\begin{figure}[t]
    \centering
    \includegraphics[width=0.48\textwidth]{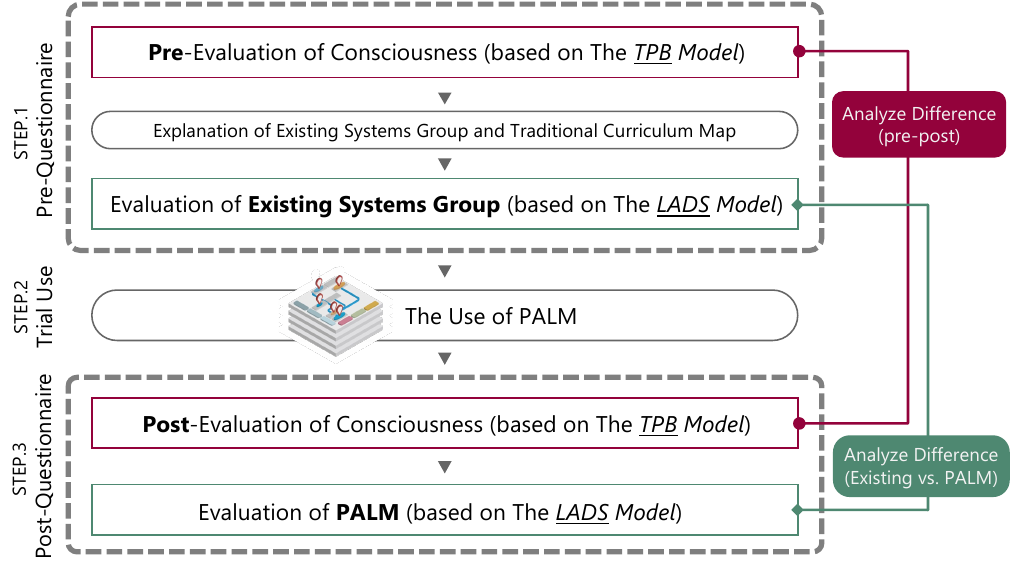} 
    \caption{Experimental procedure for system evaluation}
    \label{fig:survey_flow}
\end{figure}

To evaluate PALM, a user survey was administered to 29 participants, consisting of undergraduate students from the Department of Electrical Engineering and Computer Science at Kyushu University and master’s students who had graduated from the same department.
The survey passed an ethical review, and the participants agreed to the data use policy.
As shown in Figure~\ref{fig:survey_flow}, the survey consisted of three steps: 
(1) completing a pre-questionnaire, (2) usage of PALM, and (3) completing a post-questionnaire.
All responses were collected within two weeks.
The survey included questions focusing on \textit{system evaluation} and \textit{effectiveness evaluation}, alongside evaluation of \textit{course relevance lines}, feedback on usability and convenience, and requests for additional features. 

\subsection{Effectiveness Evaluation (RQ1)}
\label{sec:datacollection_analysys_effectiveness}
To assess the educational value of PALM, we conducted an effectiveness evaluation that compared students' behavioral intentions, planning, and reflection before and after system usage.
The survey consisted of 16 questions based on the \textbf{\textit{Theory of Planned Behavior (TPB)}}~\cite{ajzen1991theory}, a psychological model that explains how behavior is shaped by four factors: (1)~\textit{intention}, (2)~\textit{attitude toward the behavior}, (3)~\textit{subjective norm}, and (4)~\textit{perceived behavioral control}. This model was selected because it is suitable for understanding the psychological factors that influence behavioral change in the context of SRL.
Each factor was measured using 4, 6, 3, and 3 questions we prepared, respectively. 
The same items were used in both pre- and post-questionnaires, rated on a seven-point Likert scale.

By conducting statistical tests on the results of the pre- and post-questionnaires, we examined the differences in students’ attitudes toward learning and planning before and after the use of PALM. Specifically, for each respondent, the responses for each factor, which were rated on a seven-point Likert scale, were averaged to obtain a representative score. Subsequently, statistical tests were conducted based on the hypothesized causal relationships among factors derived from previous studies. First, the Shapiro-Wilk test \cite{shapiro1965test} was performed to assess whether the data followed a normal distribution. 
Next, a paired t-test (two-tailed) \cite{student1908t} was conducted to examine whether there were significant differences in the scores between the pre- and post-evaluations. 
To complement the statistical significance test, the effect size~$d_D$ was calculated using the following formula:
\begin{equation}
d_D = \frac{\overline{D} }{s_D} = \frac{\overline{(X_A - X_B)}}{\sqrt{\frac{n}{n-1} \left( s_A^2 + s_B^2 - 2s_{AB} \right) }}
\label{eq:dd}
\end{equation}
where $\overline{D}$ represents the mean difference and $s_D$ denotes the standard deviation of the differences~\cite{grissom2005effect, cohen1988statistical}. 
All statistical analyses were performed using the {\small \texttt{scipy.stats}} library in Python~\cite{virtanen2020scipy}.

\subsection{System Evaluation (RQ2)}
\label{sec:datacollection_analysys_system}

\begin{figure*}[t]
    \centering
    \includegraphics[width=1.0\textwidth]{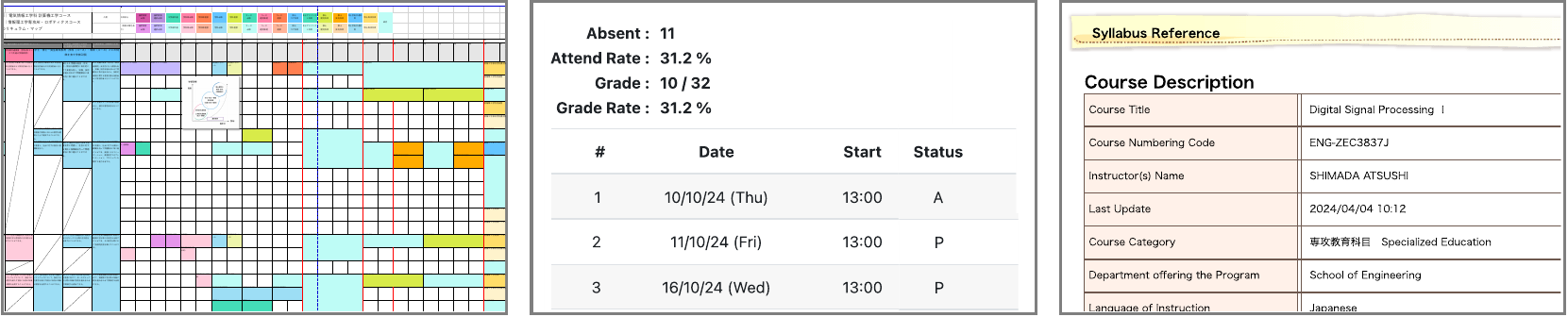} 
    \caption{Screenshots of the existing systems (traditional curriculum map $|$ attendance confirmation in LMS $|$ web syllabus)}
    \label{fig:existing_systems}
\end{figure*}

The system evaluation assessed PALM as an LA dashboard by comparing it with existing systems. 
The existing systems group included LMS, syllabus system, grade inquiry systems, traditional curriculum maps (in spreadsheet format), and course catalogs (see Figure~\ref{fig:existing_systems}).
The existing systems operate independently, and participants typically navigate between them in their regular use. 
In this evaluation, participants were asked to comprehensively consider the elements within these systems and provide an overall assessment of the existing systems group.
The survey comprised 28 items derived from the \textit{\textbf{LAD Success Questionnaire (LADS)}}~\cite{park2019factors}, a validated framework designed to assess the success of learning analytics dashboards across five factors: (1)~\textit{visual attraction}, (2)~\textit{usability}, (3)~\textit{understanding level}, (4)~\textit{perceived usefulness}, and (5)~\textit{behavioral changes}.
To support validity, we used identical questions to those which were validated in previous studies~\cite{deVreugd2024learning}.
A pre-survey was conducted to evaluate the existing systems, with items phrased as “The existing systems are/have…,” while the post-survey assessed PALM, using the phrasing “PALM is/has…” All responses used a seven-point Likert scale.
The differences in their evaluations were clarified by applying statistical tests to the evaluations of the existing system group and PALM. This followed the same procedure described in Section~\ref{sec:datacollection_analysys_effectiveness}. For each participant, the factor scores of the LADS were averaged, normality was tested using the Shapiro-Wilk test, and a paired t-test (two-tailed) was conducted.

In addition to statistical analysis, qualitative data were collected through supplementary survey questions. 
Participants were asked to evaluate whether {\footnotesize \Circled{01}}~\textit{course relevance lines} in PALM aligned with their intuition, using a five-point Likert scale. 
To obtain more in-depth insights, open-ended questions solicited feedback on various UI components, including the curriculum map, course blocks, and engagement visualizations. Participants also provided suggestions for additional features or information they would like to see incorporated into PALM, as well as potential use cases they envisioned for the system. 

\section{RESULTS}
\subsection{Effectiveness Evaluation (RQ1)}
\label{sec:result_effectiveness}
The normality test (Shapiro-Wilk test) indicated that, out of the four factors analyzed, all except \textit{attitude} met the assumption of normality. 
Subsequently, paired t-tests were performed for all of the factors, and the results are presented in Table~\ref{tbl:test_tpb}. 
For the \textit{attitude} factor, which did not meet the normality assumption, a supplementary analysis was conducted using the Wilcoxon signed-rank test \cite{wilcoxon1992individual}, a non-parametric alternative. However, because the results of both the t-test and Wilcoxon test were consistent, leading to the same conclusions regarding rejection, only the results of the t-test are reported in this study. 
The tables present the mean and standard deviation of the ratings for each factor on a seven-point scale, including the $t$-value, significance level, and effect size~$d_D$, with values above 0.8 generally considered large effects \cite{cohen1988statistical}. 

Statistical analysis revealed that, based on the TPB model comparison, post-use ratings were significantly higher than pre-use ratings for most factors, suggesting that the use of PALM had a positive impact on users’ perceptions and behavioral intentions.
Although a large statistically significant difference at the 0.1\% level was not observed for the \textit{subjective norm} factor, a significant effect was found at the 1\% level.
Effect size analysis indicated that \textit{behavioral control} showed a large effect, whereas \textit{intention}, \textit{attitude}, and \textit{subjective norm} showed only moderate effects.

\begin{table}[]
    \caption{Pre- and Post-use comparison}
    \label{tbl:test_tpb}
    \begin{tabular}{lcccc}
    \hline
    \multirow{2}{*}{\textit{\textbf{Factor (TPB)}}}      & \multicolumn{2}{c}{mean (SD)}  & \multirow{2}{*}{$t$} & \multirow{2}{*}{$|d_D|$}  \\
                                                   & pre            & post          &                      &                           \\ \hline
    \textit{\textbf{Intention}}                    & 5.1 (1.07)     & 5.7 (0.83)    & $-4.1^{***}$         & 0.77                      \\
    \textit{\textbf{Attitude}}                     & 5.0 (0.67)     & 5.6 (0.90)    & $-4.3^{***}$         & 0.80                      \\
    \textit{\textbf{Subjective norm}}              & 3.9 (1.32)     & 4.4 (1.47)    & $-3.1^{~**}$         & 0.58                      \\
    \textit{\textbf{Behavioral control}}           & 4.1 (1.10)     & 5.3 (0.87)    & $-6.5^{***}$         & 1.21                      \\ \hline
    \end{tabular}
    \\ \\
    \vspace{0.5em}
    {\footnotesize \hfill * $p < 0.05$, ** $p < 0.01$, *** $p < 0.001$ ~~~~~~}
\end{table}

\subsection{System Evaluation (RQ2)}
\label{sec:result_system}
Normality tests indicated that all five factors analyzed met the normality assumption. 
Subsequently, paired t-tests were performed for all of the factors, and the results are presented in Table~\ref{tbl:test_lads}. 
Statistical analysis revealed that PALM significantly outperformed the existing system across all evaluated factors, with large effect sizes observed.

\begin{table}[]
    \centering
    \caption{Comparison between existing systems and PALM}
    \label{tbl:test_lads}
    \begin{tabular}{lcccc}
    \hline
    \multirow{2}{*}{\textit{\textbf{Factor (LADS)}}} & \multicolumn{2}{c}{mean (SD)}  & \multirow{2}{*}{$t$}             & \multirow{2}{*}{$|d_D|$} \\
                                              & existing       & PALM          &                                  &                          \\ \hline
    \textit{\textbf{Visual attraction}}       & 3.4 (1.05)     & 6.0 (0.60)    & $-12.9^{***}$  & 2.40                    \\
    \textit{\textbf{Usability}}               & 3.7 (1.07)     & 5.9 (0.77)    & $~-9.9^{***}$  & 1.84                    \\
    \textit{\textbf{Understanding level}}     & 3.8 (1.07)     & 6.1 (0.74)    & $~-9.3^{***}$  & 1.72                    \\
    \textit{\textbf{Perceived usefulness}}    & 4.0 (1.15)     & 5.7 (0.80)    & $~-7.5^{***}$  & 1.39                    \\
    \textit{\textbf{Behavioral changes}}      & 3.7 (1.15)     & 5.3 (0.87)    & $~-6.9^{***}$  & 1.29                    \\ \hline
    \end{tabular}
    \\ \vspace{0.5em}
    {\footnotesize \hfill * $p < 0.05$, ** $p < 0.01$, *** $p < 0.001$ ~~}
\end{table}

The results of the qualitative evaluation were as follows. 
Regarding the {\footnotesize \Circled{01}}~\textit{course relevance lines}, excluding two participants who responded ``I don’t know,'' 89\% of the remaining respondents found them ``very intuitive'' or ``somewhat intuitive.'' 
Supporting comments included: ``the lines connect the recommended prerequisite courses'' and ``they reflect relationships I perceived when taking the courses.''
However, some noted the lines were ``not entirely convincing,'' citing examples such as the connection between \textit{Compiler} and \textit{Mathematics for Electrical Engineering} as counterintuitive. 
These results suggest the TF-IDF based visualization was perceived as reasonably valid, though not universally intuitive.

Open-ended responses also provided broader impressions of the system. 
Compared to the traditional spreadsheet-style curriculum map, which was criticized for ``small font size and poor readability,'' the digitalized version was praised for its ``improved text readability,'' ``clearer course relationships,'' and ``intuitive zoom functionality.''
%
Additionally, features such as {\footnotesize \Circled{02}}~\textit{visualizations of learning engagement} and {\footnotesize \Circled{03}}~\textit{comparisons with past course participants} were positively received. 
Students noted that ``the intensity of colors in course blocks facilitated intuitive comparisons'' and that ``academic grade could be linked to specific behaviors like attendance.''
However, areas for improvement were also identified, including ``excessive use of blue making elements hard to distinguish,'' ``unclear indicator meanings,'' and the need for ``the distribution and percentiles of grades.''

\addtolength{\textheight}{-0.3cm}

\section{DISCUSSION \& CONCLUSION}
\label{sec:discussion}

This study addressed the limitation of current LA research, which tends to focus on individual courses or learners.
We proposed the \textit{PAnoramic Learning Map (PALM)}, an LA dashboard that incorporates a curriculum mapping perspective, and conducted an initial evaluation for its broader educational application.

The effectiveness evaluation (RQ1) revealed a significant improvement in \textit{perceived behavioral control}, suggesting that PALM makes a key contribution as an ``information resource'' by providing previously inaccessible insights.
However, the only slight improvement in \textit{subjective norm} indicates a need to acquire and visualize more detailed peer learning data. 
The system evaluation (RQ2) showed that PALM's interface was rated favorably compared to existing systems. 
Nevertheless, open-ended feedback highlighted the necessity for supplementary guidance on its usage and for more personalized information presentation. 
In summary, while PALM requires ongoing refinement, it demonstrates clear potential for effectiveness and contribution.

Limitations of this study include the lack of support for dynamic data updates, which poses a challenge for practical implementation in large-scale settings. 
Furthermore, the potential for instructor involvement and the long-term impact on a broader student population remain under-investigated. 
Future research will therefore focus on the continuous development of the system while conducting further studies toward its scalable and practical deployment.






\section*{ACKNOWLEDGMENT}
This work was supported by JST CREST Grant Number JPMJCR22D1, JSPS KAKENHI Grant Number JP22H00551 and JP22K19835.


\bibliographystyle{IEEEtran}
\bibliography{reference}

\end{document}